# The update complexity of selection and related problems

Manoj Gupta  
Deptt. of Comp. Sc.,  
IIT Delhi, New Delhi.

Yogish Sabharwal  
IBM Research - India,  
New Delhi.

Sandeep Sen  
Deptt. of Comp. Sc.,  
IIT Delhi, New Delhi.

September 28, 2018

**Abstract**

We present a framework for computing with input data specified by intervals, representing uncertainty in the values of the input parameters. To compute a solution, the algorithm can query the input parameters that yield more refined estimates in form of sub-intervals and the objective is to minimize the number of queries. The previous approaches address the scenario where every query returns an exact value. Our framework is more general as it can deal with a wider variety of inputs and query responses and we establish interesting relationships between them that have not been investigated previously. Although some of the approaches of the previous restricted models can be adapted to the more general model, we require more sophisticated techniques for the analysis and we also obtain improved algorithms for the previous model.

We address selection problems in the generalized model and show that there exist 2-update competitive algorithms that do not depend on the lengths or distribution of the sub-intervals and hold against the worst case adversary. We also obtain similar bounds on the competitive ratio for the MST problem in graphs.

## 1 Introduction

A common scenario in many computational problems is uncertainty about the precise values of one or more parameters. Many different models have been considered in the database community for dealing with uncertain data. In one of the commonly used models, the uncertain parameters are represented by probability distributions (for a comprehensive survey, see[AY09]). In another model, the uncertain parameters are represented by interval ranges, wherein the parameter may take on any value within the specified interval (see [KT01]). In this paper, we focus on the latter model. More formally, we consider the model wherein we want to compute a function $f(x_1, x_2 \ldots x_n)$ where some (or all) $x_i$'s are not fully known. The $x_i$'s are typically known to lie in some range (interval). Any assignment of $x_i = x'_i$ consistent with the known range of $x_i$ is a *feasible realization*. The algorithm can make queries about $x_i$. This problem has been studied before [KT01, HEK[+]08]. A common assumption made in the existing literature is that the exact value of $x_i$ is returned by a single query. However, in many applications, a query about $x_i$ may only yield a more refined estimate of the $x_i$. As a matter of fact, in many such applications, it is not even possible to obtain the exact value of the parameter. As an example, consider the case of handling satellite data such as maps. Due to the large amount of data involved, the data is often stored hierarchically at different scales of resolutions. Typically the data is presented at the highest level of resolution. Depending on the area of interest, data may be retrieved for the next level of resolution for a smaller area (zoom in) by performing a query. Now consider a query to find the closest hospital. Based on the highest scale of resolution, the distances to the hospitals can be determined within a certain range of uncertainty. If the closest hospital cannot be resolved at this level, then further queries are required for certain hospitals to determine which amongst them is the closest. These queries proceed down the hierarchical scales of resolution until it is resolved which is the closest hospital.

Let us illustrate this model using the problem of finding minimum when the exact values are not known but each element is associated with a real interval $[\ell_i, r_i]$. Consider the three elements $x_1 = [3, 17], x_2 = [14, 19], x_3 = [15, 20]$. Clearly any of these can be the minimum element as these are mutually overlapping intervals. Suppose a query returns the exact value, then with three queries, we obtain the complete information and the problem is trivially solved. But the interesting question is - are three queries necessary ? Suppose our first query yields that $x_1 = 10$, then clearly we do not need to make any further queries. On the other hand, the query may yield $x_1 = 16$, so that we are forced to make further queries. In a more general situation, where a query may return a sub-interval, we may obtain $x_1 = [8, 16]$ that doesn't yield any useful information about the identity of the minimum element. On the other hand, if the query returns $[8, 10]$, then we can conclude $x_1$ to be the minimum even though we do not know the exact value of $x_1$.

It is natural to compare the number of queries made by the algorithm w.r.t. a hypothetical $OPT$ which can be thought of as a non-deterministic strategy that makes the minimum queries for any feasible realization of the input. Moreover, the algorithm must contain a certificate of correctness of the final answer, viz., that no more queries are necessary regardless of the number of unresolved parameters. This also brings up the related verification problem, i.e., given an incompletely specified problem, does it contain sufficient information for a solution to be computed (without further queries).

## 1.1 Related Previous Work

Kahan [Kah91] described a technique for maintaining data structures for online problems like flight-path collisions using predictive estimates to obtain higher efficiency. The estimates could be used to prune objects that couldn't provably affect the solution and only those *critical* objects were updated that could affect the answer. Kahan's work laid the foundations for later work on *kinetic* data structures but in his paper, he focussed on describing a framework for minimizing updates of critical objects. Kahan compared the efficiency of his data structures with respect to a non-deterministic optimal algorithm, or more specifically, the competitive ratio in the online setting. If our algorithm makes $q_S(n)$ queries for an input $S$ of size $n$, then it has competitive ratio $c$ [1] iff for some constant $\alpha > 0$,

$$q_S(n) \leq c \cdot OPT(S) + \alpha$$

where $OPT$ may be thought of as a non-deterministic algorithm (coined as *lucky* in [Kah91]) Note that $OPT$ has an unfair advantage in being able to guess the optimal sequence of queries and ensure that it can be verified in collusion with an *adversary* controlling the output of the queries.

For instance, if the given intervals are $x_1 = [2, 6], x_2 = [2, 6], x_3 = [2, 6]$, i.e., all of them are identical, $OPT$ may guess the answer to be $x_3$ and if the query yields $x_3 = 2$, then it is verified. On the other hand, an algorithm has no means of distinguishing between the $x_i$'s. Even use of randomization does not appear to provide any significant advantage in this scenario. Kahan [Kah91] tackled this issue (without acknowledging as much) by changing the problem definition to that of *reporting* **all** *values that are equal to the minimum.*

Khanna and Tan [KT01] also used the competitive ratio as a measure of efficiency of their algorithms but their parametrization didn't yield $O(1)$ bounds. Their algorithms for selection was related to the *clique number* (maximum clique size) of the input. They compare with Non-deterministic optimal and show that, no on-line algorithm can achieve a better competitive ratio than the clique number.

A somewhat different model was used by Erlebach et al.[HEK+08], who showed how to compute an *exact* minimum spanning tree for graph with interval data using minimal number of queries. The final answer is a combinatorial description (in this case a spanning tree) and not necessarily the weight of the spanning tree. Erlebach et al.[HEK+08] proved that their algorithm has competitive ratio 2 when the edge weights are initially specified as *open* intervals. One limitation of their result is the critical use of the property of open intervals which is used to weaken the advantage of $OPT$ in guessing and verifying the answer. Their results on constant competitive ratio do not hold for closed or semi-closed intervals.

A recent motivation for this line of work came from caching problems in distributed databases, (Olston and Widom [OW00]), where local cached copies are used for faster query processing where the cached values are intervals that are guaranteed to contain the actual value called the *master* value. Their work showed trade-off between the number of queries and the precision $\Delta$ of the actual answer. This model was further explored in the work of [FMP$^+$03, FMO$^+$03] that tackled fundamental problems like median-finding and shortest-paths. They distinguished between the offline (oblivious) and online (adaptive) queries including weighted versions where queries could have varying costs for different intervals. Unlike the previous work, they compared their efficiency with respect to a worst case optimal rather than a non-deterministic input-specific optimal. Therefore their results cannot be compared effectively with the previous work. Other approaches like [AH04, KZ06] minimize the worst case deviation from actual values or minimizing queries to get improved estimates of the expected solution when the distribution is known [GGM06, GM07].

---
[1] So strictly speaking, the algorithm could take exponential time but may have a bounded competitive ratio.

|     | O            | C            | OC           | P            | OP           | CP           | OCP          |
| --- | ------------ | ------------ | ------------ | ------------ | ------------ | ------------ | ------------ |
| O   | **Category-1** | (Note $\alpha$) | (Note $\alpha$) | (Note $\alpha$) | (Note $\alpha$) | (Note $\alpha$) | (Note $\alpha$) |
| C   | (Note $\alpha$) | **Category-1** | (Note $\alpha$) | (Note $\alpha$) | (Note $\alpha$) | (Note $\alpha$) | (Note $\alpha$) |
| OC  | (Note $\alpha$) | (Note $\alpha$) | **Category-1** | (Note $\alpha$) | (Note $\alpha$) | (Note $\alpha$) | (Note $\alpha$) |
| P   | trivial      | -            | -            | -            | -            | -            | -            |
| OP  | **Category-2** | (Note $\alpha$) | (Note $\alpha$) | **OP-P**     | **OP-OP**    | (Note $\alpha$) | (Note $\alpha$) |
| CP  | (Note $\alpha$) | **Category-2** | (Note $\alpha$) | **Category-3** | (Note $\alpha$) | **Category-3** | (Note $\alpha$) |
| OCP | (Note $\alpha$) | (Note $\alpha$) | **Category-2** | **Category-3** | (Note $\alpha$) | (Note $\alpha$) | **Category-3** |

Figure 1: Models for studying uncertain data problems (see note for $\alpha$ below). The allowed input types listed along the rows and the query return types listed along the columns. (The pure input point model is trivial as no queries are required).

## 2 Our contributions

In this paper, we generalize the query model in several directions. We classify models based on the types of the inputs allowed and the return type of the queries. The input may specify a combination of points (P), open intervals (I) and/or closed intervals (C). This leads to 7 variations, namely, O, C, P, OC, OP, CP and OCP. Similarly queries on intervals (open/closed) may yield points (P), open intervals (I) and/or closed intervals (C)[2]. This also leads to seven variations. These models are specified in Figure 1. We denote the models by $X$-$Y$ where $X$ denotes the type of the input allowed in the input instance and $Y$ denotes the query return types where $X$ and $Y$ can take values from O, C, P, OC, OP, CP and OCP (here the literals O, C and P correspond to open intervals, closed intervals and points respectively). Thus for instance OP-P denotes the model wherein the input can consist of open intervals as well as points and the queries can only return points.

**(Note $\alpha$):** Although there are 49 models possible, many of them are unnatural as they can lead to a change of the input type after some initial queries. The framework of such models can be covered under the framework of another suitable model. For instance, a problem under the O-P model would convert to OP-P model after a single query and is thus better studied under the OP-P model. Similarly, the OC-C model can be covered under the OC-OC model.

We categorize the valid models into 5 different categories (See Figure 1). The competitive ratios are based on this categorization of the models. *Category-1* corresponds to the models where the input and query return types are only intervals (O-O, C-C, OC-OC models). *Category-2* corresponds to the models where the input may contain points by the queries only return intervals (OP-O, CP-C, OCP-OC models). *Category-3* corresponds to the models where the input may contain closed intervals and the query may return points. The other two categories correspond to the *OP-P* and *OP-OP* models themselves.

Our main results can be summarized as follows

1. We first generalize the models to practical scenarios wherein queries may return sub-intervals as answers rather than exact values. The sub-intervals need not have any properties with respect to lengths or distributions. In other words, with further queries, we obtain increasingly refined estimates of the values until sufficient information has been obtained, i.e., the *verification* problem can be solved. We show that the *witness based approach* used in the previous models can be adapted to the models considered in this paper. More specifically, we establish interesting relationships between the various models (see Figure 2).

---
[2]We can also handle semi-closed intervals but we have avoided further classification as they don't lead to any interesting results.

2. We study the selection problem of finding the $k^{th}$ smallest value and present update competitive algorithms with different guarantees for the different models for this problem. We also study the update complexity of minimum spanning tree problem under the different models that is closely related to the extremal selection problem (finding the heaviest edge in a cycle – also called the Red rule).

3. We also show that by deviating from the witness based approach studied in prior literature, we can actually obtain improved bounds for the selection problem. These algorithms attain an *additive* overhead from optimal, that is similar to a competitive ratio of unity for some cases and are interesting in their own right.

4. Given that closed intervals have not been successfully handled in prior literature[HEK$^+$08] leading to unbounded competitive ratios, is it possible to characterize the problem more precisely? For instance, do we run into the same issues if we allow queries to return intervals? One approach for addressing issues with closed intervals is to output all the optimal solutions[Kah91]. It can be quite expensive to output all the solutions. Is there an alternate framework that addresses the issues with closed intervals without determining all the solutions.

   We show that this problem is a characteristic of models that allow closed intervals in the input and points to be returned in the queries. We extend our models to handle *closed* intervals by using the notion of lexicographically smallest solution (in case multiple solutions exist). This is a natural version in many problems where the initial ordering is important and we will show later that this has the desired effect of limiting non-deterministic guessing powers of $OPT$.

Another interesting variation could be assigning cost to a query depending on the the precision of the answer given but we have not addressed this version in this paper. There is a growing body of work that addresses the problem of computing exact answer with minimal queries [BEE$^+$06, BHKR05] and coping with more generalized queries is an important and fundamental direction of algorithmic research.

| **Problem** | **Competitive ratio** | **Models** | **Comment** | **Source** |
|---|---|---|---|---|
| Extremal selection | $OPT+1$ | OCP-P | Report all solutions | Kahan [Kah91] |
| | $OPT+1$ | OP-P | Value | this paper |
| | $2 \cdot OPT$ | Category-1,2 & OP-OP | | this paper |
| | $2 \cdot OPT$ | Category-3 | lex first | this paper |
| K-selection | $OPT+1$ | OCP-P | Report all solutions | Kahan [Kah91] |
| | $t \cdot OPT$ | CP-P | $t$ = clique no. | Khanna-Tan [KT01] |
| | $OPT+k$ | OP-P | Value, $\leq k \cdot OPT$ | this paper |
| | $2 \cdot OPT$ | Category-1 | element | this paper |
| | $2 \cdot (OPT+k)$ | OP-OP | | this paper |
| | $2 \cdot OPT$ | Category-3 | Value, lex first | this paper |
| MST | $2 \cdot OPT$ | OP-P | | Erlebach et al.[HEK$^+$08] |
| | $OPT+\mathcal{C}$ | OP-P | $\mathcal{C} \leq OPT$ $\mathcal{C}$ = no. of red rule | this paper |
| | $2 \cdot OPT$ | Category-1,2 & OP-OP | | this paper |
| | $2 \cdot OPT$ | Category-3 | lex first | this paper |

Figure 2: Known results in prior literature and our new results

## 3   Problem Definition

We consider a problem $\mathcal{P}$ where we are given an instance $P = (C, A)$ that consists of
- an ordered set of data $C = \{c_1, c_2, \ldots, c_n\}$ called a *configuration*; and

- an ordered set of data $A = \{a_1, a_2, \ldots a_n\}$ called *areas of uncertainty* such that $c_i \in a_i \; \forall i$.

The configuration $C$ is not known to us – only the areas of uncertainty, $A$, are known. As an example consider the problem, $\mathcal{P}$, of finding the index of the minimum element. An example instance is given by $P_{ex} = (C, A)$ where $C$ is the ordered set of points $C = \{3, 7, 10\}$ and $A$ is the ordered set of intervals (areas of uncertainties) $A = \{(2, 6), (5, 8), (9, 11)\}$.

We focus our discussion to problems where the input is Real data. Thus, the configuration consists of points on the Real line $\Re$, and the areas of uncertainty may be intervals on the Real line. The concepts can be extended to higher-dimensional problems.

**Verifier:** We are also given a *verifier* $V$ for the problem $\mathcal{P}$, that takes as input the areas of uncertainty, $A$ and returns whether a solution of the problem $\mathcal{P}$ can be determined from $A$ or not. For the example instance, $P_{ex}$, described above, the verifier would return false as it cannot determine a solution from the given areas of uncertainty. However, if the intervals were $A = \{(2, 5), (6, 8), (9, 11)\}$, then the verifier would return true as clearly the first interval has to contain the minimum.

**Order-Invariance:** An important characteristic of the problems we study is that the result of the verifier is only dependent on the ordering of the areas of uncertainty. More formally, consider two instances $P = (C, A)$ and $P' = (C', A')$ where $A = \{a_1, a_2, \ldots, a_n\}$ and $A' = \{a'_1, a'_2, \ldots, a'_n\}$ for the same problem $\mathcal{P}$. We say that $P$ and $P'$ are *order-equivalent* if for every pair of indices $i, j \in \{1, 2, \ldots, n\}$, it can be determined that $a_i \leq a_j$ iff it can be determined that $a'_i \leq a'_j$. We say that a problem $\mathcal{P}$ is *order-invariant* if the verifier returns the same value for any two order-equivalent configuration instances. It is easy to verify that the problems such as selection (finding minimum, finding $k^{th}$-minimum) and minimum spanning tree are order-invariant.

**Update operations:** We are allowed to perform *update* operations on the areas. Performing an update operation on area $a_i$ results in knowledge of the area to a greater degree of accuracy. More precisely, performing an update operation on $a_i$ in the instance $P = (C, A)$, where $A = \{a_1, a_2, \ldots, a_{i-1}, a_i, a_{i+1}, \ldots, a_n\}$ results in another instance $P' = (C, A')$, where $A' = \{a_1, a_2, \ldots, a_{i-1}, a'_i, a_{i+1}, \ldots, a_n\}$ such that $a'_i$ is completely contained in $a_i$. An important characteristic of the models that we consider is that the results of updates on an area are independent of updates on any other area. That is, given a multi-set $S = \{i_1, i_2, \ldots, i_k\}$ of indices of the areas, applying updates on the corresponding areas results in the same instance, irrespective of the sequence in which these updates are applied. We refer to this as the *update independence property*.

**Solution:** Our goal is to solve the problem $\mathcal{P}$ by performing minimum number of updates, i.e., perform the minimum number of updates that result in an instance for which the verifier returns true. For a problem instance $P = (C, A)$, a *solution*, $S$, is defined to be a multi-set of indices $\{i_1, i_2, \ldots, i_k\}$ such that performing updates on the areas $a_{i_1}, a_{i_2}, \ldots, a_{i_k}$ results in a problem instance $P' = (C, A')$ for which $V(A')$ returns true, i.e., a solution of the problem can be determined from $A$ without performing any more updates. In this case, we say that $S$ solves the problem instance $P$. Let $\mathcal{S}(P)$ denote the set of all such solutions. An *optimal solution* is a solution, $S \in \mathcal{S}(P)$ such that any other solution in $\mathcal{S}(P)$ has at least as many indices, i.e., $|S| \leq |S'|$ for all solutions, $S' \in \mathcal{S}(P)$. Therefore, an optimal solution corresponds to a smallest set of indices that need to be updated in order to solve the problem.

As mentioned before, the OP-P and the CP-P models have been studied before. We shall show now show that the algorithms for the OP-P model can be generalized for the many other models for problems that are order-invariant. These update competitive algorithms are based on the concept of witness sets. We discuss these concepts in Section 4; these concepts are borrowed from [BHKR05] and presented here with modifications suitable to discuss all our models. Then we discuss how to extend these algorithms to other models.

## 4 The Witness Set Framework

For a problem instance $P = (C, A)$, a set $W$ is said to be a *witness set* of $P$ if for every solution $S \in \mathcal{S}(P)$, $W \cap S \neq \phi$. Thus, no algorithm can solve $P$ without querying any area from $W$.

Suppose that we have an algorithm, `WALG`, that given any instance $P = (V, A)$ of the problem, finds a witness-set of size at most $k$. Then there exists a $k$-update competitive algorithm for the problem. The algorithm is presented in Figure 3. It simply keeps applying algorithm `WALG` to find a witness set of size at most $k$ and updates all the areas in the witness set. It keeps doing this until the problem is solved.

---

**Algorithm** `SOLVE`( Problem Instance $P$, Verifier $V$, Witness Algorithm `WALG` )
**Input**: - problem instance $P = (C, A)$,
       - a verifier algorithm $V$ for the given problem,
       - a witness algorithm `WALG` for the given problem.
**Output**: $k$-update competitive solution to problem instance $P$

Initialize solution $S = \{\}$;
If ( $V(A)$ returns *false* ) /* problem instance is not yet solved */
    $W = \texttt{WALG}(P)$;
    Update the areas in $W$ to reduce the problem instance $P$ to $P'$ ;
    $S = S \cup \texttt{SOLVE}(P', V, \texttt{WALG})$;
Endif;
Output $S$;

---

Figure 3: Algorithm to determine $k$-update competitive solution given witness algorithm

We now formally show that the solution returned by this algorithm is $k$-update competitive. Note that this result is independent of the model under consideration. The witness algorithm and verifier however are dependent on the underlying model.

**Theorem 4.1.** *The solution returned by the algorithm in Figure 3 is $k$-update competitive for the problem instance $P$.*
*Proof. See Appendix.*

**Witness Algorithms For Different Models.** Witness algorithms have been proposed for several problems under the OP-P model. We now show that the same witness algorithms can be used for various other models as well.

**Theorem 4.2.** *A witness algorithm for a problem under the OP-P model is also a witness algorithm for the same problem under the category-1, category-2 and OP-OP models (i.e., O-O, C-C, OC-OC, OP-O, CP-C, OCP-OC and OP-OP models).*
*Proof. See Appendix.*

**Corollary 4.3.** *Algorithm 3 is $k$-update competitive under the category-1, category-2 and OP-OP models with the same witness algorithms as that for the OP-P model.*
*Proof. See Appendix.*

We make an important observation here. While the reduction might seem straightforward, it is important to note many of these reductions are only one-way reduction. For instance, we can reuse the witness algorithm for the OP-P model for the OP-O model but not vice-versa. We demonstrate this later for the $k$-min selection problem, where we show that while it is possible to design a 2-update competitive algorithm under the OP-P model, it is not possible to design an algorithm that is better than $k$-update competitive under the OP-O model using witness sets.

Another important observation we make is that prior literature has shown that no algorithm can give bounded update complexity guarantees for the selection problem under the CP-P models. However, we have derived constant factor update-competitive algorithms for models involving closed intervals (i.e., the CP-C, C-C, OC-OC and OCP-OC models). This highlights the fact that the problem is not in dealing with closed intervals but rather with the combination of allowing closed intervals in the input and simultaneously allowing queries to return points for such closed intervals.

## 5 The selection problem

In an instance $P = (C, A)$ of the $k$-Min problem, $C = \{p_1, p_2, \cdots, p_n\}$ is an ordered set of points in $\Re$, and $A = \{a_1, a_2, \cdots, a_n\}$ is an ordered set of intervals on $\Re$. The nature of the intervals is determined by the model under consideration. The goal is to find the index of the $k^{th}$ smallest element in $C$.

We denote by $l_j$ and $u_j$, the lower and upper ends of the interval $a_j$ respectively. To avoid overloading of notations, we will assume that $l_j$ and $u_j$ always refer to the latest known values for the interval ranges, considering all the updates that have already been performed.

## 5.1  1-Min

In this section we look at the special case when $k = 1$, i.e., we are interested in finding the index of the smallest value interval.

**Witness Algorithm And Verifier.** We first present the witness algorithm for the OP-P model. Consider an instance $P = (C, A)$. The witness algorithm chooses the interval with the "smallest $l$-value" and the along with the interval with the next "smallest $l$-value" and returns them as the witness set. The verifier simply determines if some interval can be determined to be smaller than all the other intervals. Let $S = \{1..n\}$ denote the set of indices of the intervals. For any subset $S' \subseteq S$, we define $\text{order}_l(S')$ to be a permutation of indices in $S'$ in increasing order of the lower values of the corresponding intervals, i.e., $\text{order}_l(S') = < j_1, j_2, \cdots, j_m >$, such that $l_{j_1} \leq l_{j_2} \leq \cdots \leq l_{j_m}$. Similarly define $\text{order}_u(S') = < j_1, j_2, \cdots, j_m >$, such that $u_{j_1} \leq u_{j_2} \leq \cdots \leq u_{j_m}$.

The witness algorithm and the verifier are formally presented in Figure 4.

```
Witness Algorithm:
1. Let < p_1, p_2, ... , p_{|S|} > = order_l(S)
2. Return a_{p_1} and a_{p_2} as the witness set
```

```
Verifier:
1. Let < p_1, p_2, ... , p_{|S|} > = order_l(S)
2. If x ≤ y for all x ∈ a_{p_1} and y ∈ a_{p_j}, j ≠ 1,
       return the interval with index p_1 as the solution
   Else return false
```

Figure 4: Witness Algorithm and Verifier for 1-Min under the OP-P model

Note that an interval is declared to be the smallest interval only when no other interval can contain a smaller value. Therefore the algorithm always outputs the correct interval.

**Competitiveness.** We now show that the algorithm is 2-update competitive under the OP-P model.

**Lemma 5.1.** *The set $W = \{p_1, p_2\}$ returned by the algorithm of Figure 4 is a witness set for the 1-Min problem under the OP-P model.*
*Proof. See Appendix.*

It follows from Theorem 4.2 and Corollary 4.3 that we can derive 2-update competitive algorithms for the category-1, category-2 and OP-OP models.

**Tight Example.** We now show that the update-competitive bound of 2 is tight for all the models that allow the queries to return intervals, i.e., for the category-1, category-2 and OP-OP models (but not the OP-P model). This is demonstrated by the following example. We are given intervals $A = \{a_0, a_1, a_2, \ldots, a_n\}$ where $a_0 = (1, 5)$ and $a_j = (3, 7)$ for all $1 \leq j \leq n$. We argue that any algorithm can be forced to perform $2n$ queries while the $OPT$ can determine the interval containing the minimum with only $n$ queries. Let $S$ represent the set of intervals $A \setminus \{a_0\}$, i.e., $S = \{a_1, a_2, \ldots, a_n\}$.

Suppose that the algorithm has already performed $2n - 1$ queries. The adversary behaves as follows. For the first $n - 1$ queries on $a_0$ it returns the interval $(1 + i\varepsilon, 5)$ in the $i^{th}$ query, where $\varepsilon$ is a small value $< 1/(2n)$. For the first $n - 1$ queries on intervals from the set $S$ it returns the interval $(6, 7)$. The remaining actions of the adversary are based on whether the algorithm performs $n$ queries on $a_0$ or whether it queries $n$ intervals from $S$. Note that in performing $2n - 1$ queries, the algorithm must encounter one of these cases. These are considered in the following 2 cases:

- Case 1: The algorithm makes $n$ queries to $a_0$.
  In this case the adversary continues to return the interval $(1+i\varepsilon, 5)$ for the $i^{th}$ query on $a_0$ where $i \leq 2n-1$ and it returns the interval $(6, 7)$ for each subsequent interval queried from $S$. Note that in this case, on performing $2n-1$ queries, the algorithm could not have queried all the intervals from $S$. Therefore at the end of $2n-1$ queries, as there is overlap between interval $a_0$ and the unqueried intervals from $S$, the algorithm is forced to make $2n$ queries. The $OPT$ on the other hand can just query all the intervals in $S$. The adversary will return the interval $(6,7)$ for $OPT$ on the remaining intervals. Thus, $OPT$ is able to determine that $a_0$ contains the minimum element by just performing $n$ queries.

- Case 2: The algorithm makes $n$ queries to intervals in $S$.
  In this case, the adversary returns $(3,4)$ for the last $(n^{th})$ interval queried in $S$. For any subsequent queries to $a_0$, the adversary continues to return $(1+i\varepsilon, 5)$ for the $i^{th}$ query. Note that in this case, the adversary performs less than $n$ queries on $a_0$. Therefore at the end of $2n-1$ queries, as there is overlap between interval $a_0$ and the last queried intervals from $S$, the algorithm is forced to make $2n$ queries. The $OPT$ on the other hand can just query all the intervals in $a_0$. The adversary will return the value $(2,3)$ for $OPT$ on its $n^{th}$ query to $a_0$ (recall that in this case the algorithm did not perform $n$ queries on $a_0$). Thus, $OPT$ is able to determine that $a_0$ contains the minimum element by just performing $n$ queries.

It is surprising that though this tight example demonstrates that we cannot obtain better than 2-update competitive algorithms for these models, it is possible to obtain a 1-update competitive algorithm for the OP-P model; however, this is obtained by an approach different from the Witness Set framework. This is discussed in more detail in Section 6.

## 5.2 $K$-Min

We now generalize the 1-min algorithm presented above to the $k^{th}$-min problem, but under the O-O model. We later discuss issues related to handling points under the OP-P model.

**Witness Algorithm And Verifier.** We now present a witness algorithm and verifier for this problem under the O-O model.

```
Witness Algorithm:
1. Let < p_1, p_2, ··· , p_n > = order_l(S)
2. Let S' = {p_1, .., p_{k-1}}
3. If x ≤ y ∀ x ∈ a_i, i ∈ S' and ∀ y ∈ S \ S'
       return the witness set of 1-Min algorithm
4. Else
       let < q_1, q_2, ··· , q_{|S'|} > = order_u(S')
       return a_{p_k} and a_{q_1} as the witness set
```

```
Verifier:
1. Let < p_1, p_2, ··· , p_n > = order_l(S)
2. Let S' = {p_1, .., p_{k-1}}
3. If (x ≤ y ∀ x ∈ a_i, i ∈ S' and ∀ y ∈ a_{p_k}) and
       (x ≥ y ∀ x ∈ a_i, i ∈ S \ (S' ∪ a_{p_k}) and ∀ y ∈ a_{p_k})
       return a_{p_k}
   else return false
```

Figure 5: Witness and Verifier Algorithm for K-Min under the O-O model

We say intervals $a_i$ and $a_j$ are disjoint if $\forall x \in a_i, y \in a_j, x \leq y$ or vice-verse. The witness algorithm checks if the first $k-1$ interval are *disjoint* with the last $n-k+1$ interval. If that is the case, it returns the witness set of the 1-Min algorithm. Else it chooses $a_{p_k}$ and an interval from $S'$ with largest $u$ value($a_{q_1}$) as the witness set.

The $verifier$ takes the first $k-1$ intervals($S'$) depending on their $l$ values. The $verifier$ checks if these $k-1$ intervals are *disjoint* from the $a_{p_k}$. Then it takes the last $n-k$ intervals($S \setminus (S' \cup a_{p_k})$) and checks if all of them disjoint with $a_{p_k}$. If both the condition holds, it returns $a_{p_k}$ else it returns false.

**Competitiveness.** We now show that the algorithm is 2-update competitive for the O-O model. It follows using proofs similar to Theorem 4.2 and Corollary 4.3 that we can derive 2-update competitive algorithms for the other category-1 models.

**Lemma 5.2.** *The witness set $W$ returned by the algorithm of Figure 5 is a witness set for the k-Min problem under the O-O model.*
*Proof. See Appendix.*

**Tight Example.** It is not difficult to construct examples similar to that discussed for the 1-Min algorithm to show that the update-competitive bound of 2 is tight under the category-1 models.

It is interesting to note here that while a 2-update competitive algorithm can be designed for the $k$-min problem under the category-1 models, no algorithm can be better than $k$-update competitive for this problem under models that allow points, i.e., the category-2 and OP-P models. This is illustrated by the following example[3]. Suppose we have $2k$ areas of which $k$ are open intervals of the form $(0,5)$ and $k$ are fixed points of the value 3. For the first $k-1$ intervals queried by any algorithm, the adversary returns 1 and for the $k^{th}$ interval, the adversary returns 4 (or interval (3.5,4.5) as the case may be), thereby forcing $k$ queries. However, OPT only needs to update the interval with value 4 and can thereafter return any of the $k$ fixed points of value 3 as the $k^{th}$ smallest.

However, in the next section we show that it is possible to design algorithms for the $k$-Min problem under these models that allow for points, obtaining update competitive bounds with additive factor $k$ (i.e., the algorithm performs $k$ more updates than $OPT$). This however is achieved by bypassing the Witness set framework.

## 6 Bypassing the Witness Set framework

While the witness set framework, studied in prior literature, provides a general method for solving problems with data uncertainty under the update complexity models, it has its limitations. We demonstrate this by presenting algorithms that require to perform only $k$ more queries than OPT for the $k^{th}$-Min selection problem. Note that, for the 1-Min problem this implies a 1-update competitive algorithm, as only one query more than OPT is required to be performed.

### 6.1 1-Min

Consider the following algorithm. We note here that the set of intervals returned by the "witness" algorithm

```
``Witness'' Algorithm:
1.  Let < p_1, p_2, ··· , p_{|S|} > = order_l(S)
2.  Let A = {a_{p_1}} and B = {p_2, ··· , p_{|S|}}
3.  Return interval in A.
```

```
Verifier:
1.  Let < p_1, p_2, ··· , p_{|S|} > = order_l(S)
2.  If x ≤ y for all x ∈ a_{p_1} and y ∈ a_{p_j}, j ≠ 1,
       return the interval with index p_1 as
       the solution
    Else return false
```

Figure 6: "Witness" Algorithm and Verifier for 1-Min under the OP-P model

is not a true witness set. However, we stick to the terminology for the sake of consistency. The algorithm remains the same, it updates the intervals returned by the "witness" algorithm until we obtain a solution.

**Lemma 6.1.** *Let $c_{OPT}$ be the total number of queries made by OPT to find 1-Min, then total number of queries made by algorithm in Figure 6 is at most $c_{OPT} + 1$ in the OP-P model.*
*Proof. See Appendix.*

Note that this simple algorithm for 1-Min in OP-P model fails for the OP-O model. Consider the following example. Let there be two intervals $I_1 = (2, 20)$ and $I_2 = (19, 21)$ Suppose at the $i^{th}$ query of $I_1$, we get a new interval $(d_i, 20)$, where $d_i < 19$, so $I_1$ and $I_2$ will always intersect if we just query $I_1$. The algorithm in Figure 6 always queries $I_1$, so it takes huge number of queries to find 1-Min. But if we just query $I_2$, it returns a subinterval (20.5,21). This is what OPT does and uses just one query to find the answer.

### 6.2 $k$-Min

Consider the algorithm in Figure 7 for k-selection in the OP-P model which generalizes the result of the algorithm in Figure 6.

---
[3]This was pointed out by an anonymous reviewer of a previous version

```
''Witness'' Algorithm:                                  Verifier:
1. Let < p_1, p_2, ···, p_n > = order_l(S)               1. Let < p_1, p_2, ···, p_n > = order_l(S)
2. Let S' = {p_1, .., p_k}                               2. Let S' = {p_1, .., p_{k-1}}
3. let < q_1, q_2, ···, q_k > = order_u(S')              3. If (x ≤ y ∀ x ∈ a_i, i ∈ S' and ∀ y ∈ a_{p_k}) and
   Let S'_max = a_{q_k}. Query S'_max.                      (x ≥ y ∀ x ∈ a_i, i ∈ S \ (S' ∪ a_{p_k}) and ∀ y ∈ a_{p_k})
4. If x ≤ y ∀ x ∈ a_i, i ∈ S' and ∀ y ∈ S \ S'              return a_{p_k}
   return the "witness set" of the                       else return false
   1-Max algorithm of S' (of Figure 4).
```

Figure 7: Witness and Verifier Algorithm for K-Min under the OP-P model

**Lemma 6.2.** *The algorithm of Figure 7 uses atmost $c_{OPT} + \min\{k, n-k\}$ queries where $c_{OPT}$ is the minimum number of queries required by the OPT.*
*Proof. See Appendix.*

Now let us consider the OP-OP model. Note that since we have $2 \cdot OPT$ algorithms for the OP-O model and an $OPT + k$ algorithm for the OP-P model, we can derive a $2 \cdot (OPT + k)$ algorithm for the OP-OP model by combining these 2 algorithms. This is done by alternating the witness algorithms of the two models. This ensures that we only need to perform at most twice the number of queries performed by the algorithms of either of the two models.

## 7 Closed intervals with point returning queries

As discussed above, the competitive ratio is unbounded for the special cases where the input allows for closed intervals and queries may return points (i.e., the category-3 models). For instance consider the problem of finding the index of the minimum element. Further, consider the problem instance $P = (C, A)$ where $a_i = [1, 3]$ for all $1 \leq i \leq n$. The adversary in this case acts as follows; for each of our queries except the last, it returns 2. Finally, for our last query, say on interval $a_k$, it returns 1. On the other hand, $OPT$ directly queries interval $a_k$ and obtains the optimal solution. This results in an unbounded competitive ratio.

The primary reason for this anomaly is the possibility of existence of multiple optimal solutions. In such cases, the adversary is able to get away with few queries by just querying the necessary intervals that reveal one of the optimal solutions. For any algorithm on the other hand, it is not able to distinguish from the areas of uncertainty (as shown above) which are the necessary intervals to query to reveal the optimal solution.

One of the ways that has been suggested in prior literature to deal with this special case is to require all the optimal solutions to be output. However, it can be quite expensive to output all these solutions. This raises the question of whether other reasonable conditions can be laid on the structure of the required output that are not so expensive but reasonable. We now consider such a condition, which we call the *lexicographic condition*, for which we show that this special can be handled. Recall that the sets $C$ and $A$ that define a problem instance are ordered sets. Thus, the set of indices that define a solution can be considered as a string (called *solution string*) defined as follows: the length of the string is $n$ and the $i^{th}$ element of the string is set to 1 if it defines the solution and 0 otherwise. In the lexicographic setting, amongst all the optimal solutions, we are interested in finding the solution for which the solution string has the smallest lexicographic ordering.

Now consider again the example above. Note that, even though $OPT$ queries $a_k$ and determines a solution with optimal solution value, it cannot terminate without making further queries as it cannot decide whether or not there exists another solution with the same value but a smaller lexicographic ordering.

We note that new witness algorithms may require to be developed for the lexicographic variants of the problems. However, we show by case of examples that these are not very different from the corresponding witness algorithms for the original problems.

It can be shown that once a witness algorithm is developed for a lexicographic variant of the problem under the CP-P model, the same witness algorithm can be extended to other models along the same lines as discussed in Section 4.

Now let us consider the lexicographic variant of the 1-Min problem. In order to obtain the witness algorithm for the lexicographic variant for the category-3 models, the notion of ordering of intervals, $\text{order}_l(.)$, needs to be extended to incorporate lexicographic ordering and closed intervals. As before, for any subset $S' \subseteq S$, we define $\text{order}_l(S')$ to be a permutation of indices in $S'$ in increasing order of the lower values of the corresponding intervals, i.e., $\text{order}_l(S') = <j_1, j_2, \cdots, j_m>$, such that $l_{j_1} \leq l_{j_2} \leq \cdots \leq l_{j_m}$. When comparing two intervals with the same $l$-values, say $l_j$ and $l_{j'}$, ties are resolved as follows: If $a_j$ contains a point $x$ such that $x < y$ for all $y \in a_{j'}$, then $j$ precedes $j'$ in the ordering; similarly if $a_{j'}$ contains such a point, then $j'$ precedes $j$; and if neither can be established, then the lexicographically smaller index precedes the larger one in the ordering. Thus, if one of the intervals, say $a_j$, is open from the left and another interval, say $a_{j'}$, is either closed from the left or a point, then $j'$ precedes $j$ in the ordering; in all other cases, the lexicographic smaller of $j$ and $j'$ precedes the other in the ordering.

The witness algorithm and verifier are formally presented in Figure 8. Note that the verifier is also modified so that it can check that the minimum interval can be determined or not based on the lexicographic ordering.

```
Witness Algorithm:
1. Let < p_1, p_2, ..., p_|S| > = order_l(S)
2. Return a_{p_1} and a_{p_2} as the witness set
```

```
Verifier:
1. Let < p_1, p_2, ..., p_|S| > = order_l(S)
2. If (x ≤ y ∀ x ∈ a_{p_1} and y ∈ a_{p_j}, p_j > p_1) and
       (x < y ∀ x ∈ a_{p_1} and y ∈ a_{p_j}, p_j < p_1),
         return the interval with index p_1 as the solution
   Else return false
```

Figure 8: Witness Algorithm for 1-Min under the CP-P model

The proof of update competitiveness is similar to the case for the original problem.

**Lemma 7.1.** *The set $W = \{p_1, p_2\}$ returned by the algorithm of Figure 8 is a witness set for the lexicographic 1-Min problem under the CP-P model.*
*Proof. See Appendix.*

The fact that no algorithm can be better than 2-update competitive for the 1-Min problem under the CP-P model follows from the same reasoning as for the OP-P model.

We can extend this 2-update competitive algorithm for the other category-3 models using techniques similar to that in Section 4.

Finally, we can design 2-update competitive algorithms for the $k$-min version as well under these models by using similar techniques.

## 8 Minimum Spanning Tree

In the Lexicographic MST problem, we are given a graph $G = (V, E)$. The edge lengths are specified with uncertainty. Let $E = \{e_1, e_2, \ldots, e_n\}$ be the ordered set of edges. Then the ordered set $C = \{v_1, v_2, \cdots, v_n\}$ denotes the values of the edge lengths and the ordered set $A = \{a_1, a_2, \cdots, a_n\}$ denotes the intervals within which the edge lengths are known to lie. The goal is to find the lexicographically smallest MST under the category-3 models.

A 2-update competitive algorithm for the MST problem was given by [HEK+08] under the OP-P model. By applying Theorems 4.2 and Corollary 4.3, we conclude that it is 2-update competitive for the Category-1,2 and OP-OP models as well. The Lexicographic MST problem can be solved under the Category-3 models with few changes to the algorithm described in [HEK+08] (these changes are outlined in Appendix A). This gives us the following result.

**Theorem 8.1.** *There exists a 2-update competitive algorithm for the Lexicographic MST problem under the Category-3 models.*

*Remark:* It may be noted that the algorithm described in [HEK+08] in conjunction with Lemma 6.1 can be used to derive an $OPT + \mathcal{C}$ update competitive algorithm for the MST problem under the OP-OP model where $\mathcal{C}$ is the number of red-rules applied by the optimal algorithm. Note that $\mathcal{C}$ can be much less than $OPT$.

## 9 Conclusion

We extended the one-shot query model to the more general situation where a query can return arbitrary sub-intervals as answers and established strong relationships between these models. Many of the previous results in the restricted model can be generalized based on this relationship that simplifies the task of designing algorithms for the more general model. This is far from obvious as the sub-interval query model presents some obvious challenges because the uncertainty (in the values of any parameter) can take an arbitrary number of steps to be resolved and can be controlled by an adversary. One drawback of this approach is that the actual algorithmic complexity is overlooked and we only focus on the competitive ratio which is justified on the basis of very high cost of a query. For future work, the algorithmic complexity needs to be incorporated in a meaningful way.

## Appendix A. Sketch of changes for Lexicographic MST

The following changes are required to the algorithm of [HEK+08]. Here we use the notation $U_x$ for $u_x$ and $L_x$ for $l_x$ to remain consistent with [HEK+08].

1. The main change involves modifying the comparison operator. We modify the comparison operator defined on the intervals as follows: Let $x$ be the $l$-value or $u$-value of some interval, i.e., $x = l_e$ or $x = u_e$ for some interval $e$. Similarly, let $y$ be the $l$-value or $u$-value of some interval, i.e., $y = l_f$ or $y = u_f$ for some interval $f \neq e$. We say that $x \prec y$ if $x < y$ or $x = y$ and $e$ is lexicographically smaller than $f$.

2. An edge $e$ of a cycle $C$ is said to be always maximal if $U_c \prec L_e$ for all $c \in C - \{e\}$. Note that the only change introduced in this definition is in replacing the comparison operator.

3. We similarly modify the notion of comparing two edges $e$ and $f$ based on the comparison operator as follows. We say that $e \prec f$ if $L_e \prec L_f$. While indexing the edges in the algorithm, the edges are considered in the order defined by $\prec$ above.

4. The witness set is determined as follows. Once a cycle $C$ is detected, if it contains an always maximal edge, that edge is deleted. Otherwise let $f \in C$ such that $U_f = \max\{U_c | c \in C\}$ where max is based on the new $\prec$ operator. Further let $g \in C - \{f\}$ such that $L_f \prec U_g$. Then $f$ and $g$ form the witness set.

Theorem 8.1 can be proved with these changes along the same lines as presented in [HEK$^+$08].

# Appendix B. Proofs for the Witness Set Framework

## 9.1 Proof of Theorem 4.1

We first prove a claim that will be required in the proof of the above result.

**Claim 9.1.** *Suppose that we are given a problem instance $P = (C, A)$. Further, suppose that we know that an optimal solution, $S_o$ for $P$ contains an index $i$, i.e., $S_o$ queries the area $a_i$. Let $P' = (C, A')$ be the problem instance reduced from $P$ on querying area $a_i$. Then $S'_o = S_o \setminus \{i\}$ is an optimal solution for $P'$. Here the operation $\setminus$ on the multiset $S_o$ removes only one instance of $i$ from it in case there are multiple instances.*
*Proof. See Appendix.*

*Proof.* Recall that the update independence property implies that irrespective of the order in which the updates are applied, applying all the updates in $S_o$ solves the problem $P$. Therefore, clearly $S'_o$ solves the problem instance $P'$. In order to argue that this is an optimal solution, all we need to show is that there does not exist a solution of smaller size. Suppose otherwise. Then there exists a solution $S$ of size smaller than $S'_o$ that solves $P'$. But then, $S' = S \cup \{i\}$ solves $P$ which contradicts the fact that $S_o$ is an optimal solution of $P$. □ □

We now present the proof of Theorem 4.1.

*Proof.* The proof is by induction on the size of an optimal solution on instance $P$. For the base case, consider a problem instance $P$ for which any optimal solution has size 1. Let $W$ be a witness set returned by algorithm WALG. Clearly, $W$ is $k$-update competitive by definition.

Now suppose that the claim holds for any problem instance $P$ having optimum solution of size $i$ or less. Consider a problem instance $P$ for which any optimum solution has size $i + 1$. Let $W$ be a witness set of size $\leq k$ returned by WALG. Let the instance $P$ be reduced to instance $P'$ on applying updates on the areas in $W$. By Claim 9.1, any optimal solution on $P'$ has size $\leq i$. By induction, the algorithm determines a $k$-update competitive solution $S'$ for $P'$. Hence $|S' \cup W| \leq k(i+1)$, and thus $S' \cup W$ is a $k$-update competitive solution for $P$. □ □

## 9.2 Proof of Theorem 4.2

*Proof.* We formally prove this for the CP-C model. The proofs for the other models follow similarly; we point out the changes required.

Let $P$ be any instance of the given problem under the CP-C model. Let $P'$ be obtained from $P$ by modifying the configuration and areas of uncertainty as follows; (i) All the closed intervals are replaced with open intervals; and (ii) The configuration is suitably modified in order to ensure that the configuration points are always contained in the corresponding areas of uncertainty – this is explained in more detail later. Let $W$ be any witness set for $P'$ under the OP-P model. We need to show that $W$ is also a witness set for $P$ under the CP-C model. Suppose this is not so, i.e., $W$ is not a witness set for $P$ under the CP-C model. We will then argue that there exists a set of queries excluding $W$ that when applied to $P'$ under the OP-P model can result in an instance for which the verifier returns true; this implies that $W$ is not a witness set for $P'$ under the OP-P model leading to a contradiction. Hence our supposition is incorrect and $W$ must be a witness set for $P$ as well.

It remains to find a possible set of queries excluding $W$ and query outcomes that when applied to $P'$ under the OP-P model results in an instance for which the verifier returns true under the assumption that $W$ is not a witness set for $P$ under the CP-C model. Considering this assumption, there exists a solution $S = \{i_1, i_2, \ldots, i_k\}$ for $P$ under the CP-C model that does not contain the index for any area in $W$. Let $P_1, P2, \ldots, P_k$ be the sequence of instances obtained on applying the updates in $S$ where $P_t$ is obtained from $P_{t-1}$ on applying the update on $a_{i_t}$ for $1 \leq t \leq k$. For any interval (not a point) $a_j$ in $P_k$ (the final configuration in the sequence above) let $l_j, u_j$ denote the interval end points. Let $\varepsilon = \min\{u_j - l_j | a_j \ a_j \text{ is an interval in } P_k\}$, i.e., $\varepsilon$ is the minimum length of any interval in $P_k$.

As mentioned earlier, the configuration points in $P'$ are also suitably modified in order to ensure that they are always contained in the corresponding areas of uncertainty. This is done by setting a configuration point $c_j$ to $c_j + \varepsilon/10$ if $c_j = l_j$ in $P_k$, and setting it to $c_j - \varepsilon/10$ if $c_j = u_j$ in $P_k$. This ensures that no configuration point coincides with the interval end-points; this will allow us to replace closed intervals with open intervals. Moreover, the modified configuration points are consistent with all the query outputs.

Now, consider the case where the same sequence of updates in $S$ is applied to $P'$ under the OP-P model. A possible sequence of outcomes is $P_1', P_2', \ldots, P_k'$ wherein $P_t'$ is the same as $P_t$ with all closed intervals replaced by open intervals ($P_t'$ is obtained from $P_{t-1}'$ on applying the update on $a_{i_t}'$ for $1 \leq t < k$). Now note that since $S$ solves $P$, the verifier returns true for $P_k$. However, $P_k$ and $P_k'$ are order-equivalent and since the problem is order-invariant, the verifier must return true for $P_k'$ under the OP-P model as well. This implies that $W$ is not a witness set for $P'$ under the OP-P model leading to the required contradiction.

We can similarly show that the witness set for the OP-P model can be reused for a variety of other models, thereby resulting in comparable update-competitive algorithms. The proofs are similar to that of Theorem 4.2 above; the only difference is in the way the the instances $P_1', \ldots, P_k'$ for the OP-P model are constructed from the instances $P_1, \ldots, P_k$ for the new model.

For the C-C model, OC-OC model and the OCP-OC model, the instance $P_t'$ is obtained from instance $P_t$ by replacing the closed intervals in $P_t$ with corresponding open intervals and modifying the configuration points as described in the proof above.

For the O-O model, and the OP-O model, the instance $P_t'$ is obtained from instance $P_t$ by replacing the intervals in $P_t$ corresponding to the areas of uncertainty having indices in the set $\{i_1, i_2, \ldots, i_t\}$ with the corresponding configuration points $c_{i_1}, c_{i_2}, \ldots, c_{i_t}$. Note that in this case, it does not make sense to query the interval on the same index more than once, therefore the number of queries can be reduced.

This completes the proof of the Theorem. □ □

## Proof of Corollary 4.3

*Proof.* Consider the CP-C model. By Theorem 4.2, we know that the witness algorithm for the OP-P model is also a witness algorithm for the CP-C model. Moreover, the verifier for the OP-P model is also a verifier for the CP-C model as the problem considered is order-invariant. The proof for the other models follows similarly. □ □

# Appendix C. Proofs for the Selection Problem

## Proof of Lemma 5.1

*Proof.* The proof is by contradiction. Suppose $OPT$ updates neither $a_{p_1}$ nor $a_{p_2}$. Let the index of the interval returned by $OPT$ as the answer be $a_q$. We consider the 2 cases:

- $a_q = a_{p_1}$: As the witness algorithm is invoked only when the verifier returns false, by examining the condition in Step 2 of the verifier (which must have failed for the current instance), we conclude that $\exists$ $x \in a_{p_1}$ and $y \in a_{p_j}, j \neq 1$ such that $y < x$. Thus $OPT$ has not fully demonstrated that $a_{p_1}$ contains the point which is minimum as $a_{p_2}$ could be made to contain the minimum point.

- $a_q \neq a_{p_1}$: By the definition of $\text{order}_l(.)$ applied in Step 1 of the witness algorithm and by examining the condition in Step 2 of the verifier, we conclude that $l_{p_1} \leq l_{p_q}$. Thus, $OPT$ has not demonstrated that $a_{p_q}$ contains the point which is minimum as $a_{p_1}$ could be made to contain the minimum point. □

□

## Proof of Lemma 5.2

*Proof.* The proof is again by contradiction. There are two cases:

- *The witness set returned is the witness set of the 1-Min Algorithm, $W = \{a_{p_k}, a_{p_{k+1}}\}$:*

  Since the first $k-1$ intervals are *disjoint* with the rest of the intervals, the problem of finding the $k^{th}$ minimum interval becomes the problem of finding 1-Min in $S \setminus S'$. Using Lemma 6.1, $W$ is a valid witness set.

- $W = \{a_{p_k}, a_{q_1}\}$:

  Suppose $OPT$ updates neither $a_{p_k}$ nor $a_{q_1}$. Let the index returned by $OPT$ be $a_j$. So $a_j$ has to be *disjoint* with all the other intervals. Since the witness algorithm was called only because the verifier returned false, so by examining the condition of step 3 of the verifier, we infer that $\exists a_{p_i}$ with $1 \leq i \leq k-1$ such that $\exists x \in a_{p_i}, y \in a_{p_k}$ for which $x > y$. So $a_{p_i}$ and $a_{p_k}$ are not *disjoint*. If such $a_{p_i}$ exists, then by the definition of $a_{q_1}$, we see that $a_{q_1}$ and $a_{p_k}$ are also not *disjoint*. So the solution returned by $OPT$ cannot be $a_{p_k}$ and $a_{q_1}$ as they both are not *disjoint*. As $a_j$ must be disjoint, we consider following cases:

    - $u_j \leq l_{p_k}$ and $u_j \leq l_{q_1}$: Initially there were less than $k-2$ intervals with $l$ values $\leq l_{q_1}$. Since $a_{q_1}$ is not updated, any update of other intervals cannot increase the number of intervals with $l$ values $\leq l_{q_1}$. Since $u_j \leq l_{p_k}$, the number of intervals with $l$ values $\leq l_j$ is less than $k-3$. So $a_j$ cannot be the $k^{th}$ minimum interval.

    - $l_j \geq u_{p_k}$ and $l_j \geq u_{q_1}$: Initially there are $k-2$ intervals with $u$ values $\leq u_{q_1}$. Since $a_{q_1}$ is not updated, any update of other intervals is not going to decrease the number of such intervals. These intervals together with $q_1$ and $p_k$ have $u$ values $\leq l_j$. So there are $k$ intervals with $u$ values $\leq l_j$. So $a_j$ cannot be the $k^{th}$ minimum interval. □

□

# Appendix D. Proofs for Bypassing the Witness set Framework

## Proof of Lemma 6.1

*Proof.* Assume for contradiction that we have queried $c_{OPT} + j$ intervals where $j \geq 2$. Let $a_1$ and $a_2$ be any two intervals that algorithm in Figure 6 has queried but OPT has not queried such that $l_1 \leq l_2$. Since OPT did not query $a_1$, we conclude that $a_1$ is the interval which contains the minimum. Also since the algorithm in Figure 6 queried $a_2$, $\exists\ x \in a_1$ and $y \in a_2$ such that $y < x$. But we have assumed that OPT does not query $a_2$, so OPT cannot demonstrate that $a_1$ contains the point which is minimum. So we get a contradiction. □

□

**Proof of Lemma 6.2**

*Proof.* Let $1 < k \leq n-k$ - the other case can be argued similarly and $k=1$ is addressed by the algorithm in Figure 6. If $S'_{\max}$ is not queried by OPT then $S'_{\max}$ has rank $\leq k$. $S'_{\max}$ cannot have rank $> k$ by definition of $S'$. Indeed, if $S'_{\max}$ has rank $> k$, then there must be at least $k$ points to the left of $S'_{\max}$ that violates the definition of $S'$. If $S'_{\max}$ has rank $\leq k$ then atmost $k-1$ such intervals can remain unqueried, otherwise the rank of the element returned cannot provably be $k$. (If $S'_{\max}$ has rank $= k$, then the OPT must query all except one, which is $\leq k-1$ for $k > 1$). For the second phase, to find out the maximum among $S'$, the algorithm of Figure 6 needs at most $c_{OPT}^{\max}+1$ queries. So, overall, our algorithm makes at most $k-1+1 = k$ queries more than the OPT. □

## Appendix E. Proofs for Closed intervals with point returning queries

**Proof of Lemma 7.1**

*Proof.* The proof is again by contradiction. Suppose $OPT$ updates neither $a_{p_1}$ nor $a_{p_2}$. Let the index of the interval returned by $OPT$ as the answer be $a_q$. We consider the 2 cases:

- $a_q = a_{p_1}$: As the witness algorithm is invoked only when the verifier returns false, by examining the condition in Step 2 of the verifier (which must have failed for the current instance), we conclude that either (i) $\exists\, x \in a_{p_1}$ and $y \in a_{p_j}, j \neq 1$ such that $y < x$; or (ii) $\exists\, x \in a_{p_1}$ and $y \in a_{p_j}$ such that $y = x$ and $p_2 < p_1$. In either case, we observe that $OPT$ has not fully demonstrated that $a_{p_1}$ contains the point which is minimum as $a_{p_2}$ could be made to contain the minimum point.

- $a_q \neq a_{p_1}$: By the definition of $\texttt{order}_l(.)$ applied in Step 1 of the witness algorithm and examining the condition in Step 2 of the verifier, we conclude that $l_{p_1} \leq l_{p_q}$. Thus, $OPT$ has not demonstrated that $a_{p_q}$ contains the point which is minimum as $a_{p_1}$ could be made to contain the minimum point. □

□